\documentclass[%
 reprint, 
 amsmath,amssymb,
 aps,
 prb,
]{revtex4-1}  

\usepackage{graphicx}
\usepackage{dcolumn}
\usepackage{bm}
\usepackage{epstopdf}

\setcitestyle{super}
\newcommand*{\citen}[1]{%
  \begingroup
    \romannumeral-`\x 
    \setcitestyle{numbers}%
    \cite{#1}%
  \endgroup
}
\begin{document}

\preprint{APS/123-QED}

\title{Lattice dynamics and ferroelectric properties of nitride perovskite LaWN$_3$}

\author{Yue-Wen Fang$^{1,2,\ddag}$}
\author{Craig A.J. Fisher$^2$}
\author{Akihide Kuwabara$^2$}
\author{Xin-Wei Shen$^1$}
\author{Takafumi Ogawa$^2$}
\author{Hiroki Moriwake$^{2,\dag}$}
\author{Rong Huang$^{1,2}$}
\author{Chun-Gang Duan$^{1,3,}$}%
\email{cgduan@clpm.ecnu.edu.cn,
$\dag$ moriwake@jfcc.or.jp,
$\ddag$ fyuewen@gmail.com}

\affiliation{%
 $^1$Key Laboratory of Polar Materials and Devices,\!
 Ministry of Education,\! Department of Electronic Engineering,\!
  East China Normal University, Shanghai 200241, China\\
 $^2$Nanostructures Research Laboratory, Japan Fine Ceramics Center, Nagoya 456-8587, Japan\\
 $^3$Collaborative Innovation Center of Extreme Optics, Shanxi University,
 Taiyuan, Shanxi 030006, China\\}



%

\date{\today}

\begin{abstract}
Using first-principles calculations we examine the crystal structures and phase transitions of nitride perovskite LaWN$_3$. Lattice dynamics calculations indicate that the ground-state structure belongs to space group $R3c$. Two competitive phase transition pathways are identified which are characterized by symmetry-adapted distortion modes. The results suggest that $R3c$ LaWN$_3$ should be an excellent ferroelectric semiconductor: its large spontaneous polarization of around 61 $\mu$C/cm$^2$ is comparable to that of PbTiO$_3$, and its band gap is about 1.72 eV. Ferroelectricity is found to result from the \emph{B}-site instability driven by hybridization between W-5$d$ and N-2$p$ orbitals. These properties make LaWN$_3$ an attractive candidate material for use in ferroelectric memory devices and photovoltaic cells.
\end{abstract}

\maketitle


\section{\label{sec:level1} INTRODUCTION}
Perovskites ${ABX}$$_3$, in which 12-fold coordinated $A$-site cations each sit between eight corner-sharing ${BX}$$_6$ octahedra, exhibit a vast variety of physical, chemical, electrical and magnetic properties, making them of enormous interest from both a scientific and technological point of view.\cite{Gruverman2015,Pennycook-science2015,Ghosez2015,Rabe2016PRX,LiuHJ2016} For example, ferroelectric oxide perovskites ${AB}$O$_3$, long established as tunable capacitors, ultrasound generators, infrared sensors, and fuel injectors, are now being developed for use in data storage devices\cite{Garcia2014NatComm,Lee2012} and photovoltaic cells.\cite{Photo-review2015,Nechache-SolarNatPhot2015}

At present, commercial ferroelectric random-access memory uses lead zirconate titanate (PZT),\cite{Juan2009} which is potentially hazardous because of the toxicity of lead.\cite{Panda2009} There is thus a strong incentive to develop alternative materials exhibiting excellent ferroelectric or photoferroic properties that are non-toxic and environmentally benign.

A number of studies in recent years have used computational screening methods to search for novel candidate ferroelectric materials.\cite{MRS-2016,Ye2016,Xie-scirep-2015,Armiento-2011PRB} Of the many new systems identified, perovskites with the LiNbO$_3$-type structure that crystallize in space group $R3c$ are particularly interesting because the driving mechanisms behind their ferroelectricity are very different to those in the extensively studied BaTiO$_3$ and PbTiO$_3$ systems.\cite{Belik2016,Wang2015PRL,HAO-R3C2012PRB,Zhang2010,Duan2004NaCdF3,CohenPRB1996} For example, LiNbO$_3$-type FeTiO$_3$ has been predicted to exhibit a spontaneous polarization, despite its lack of lone-pair electrons, that simultaneously induces a weak ferromagnetism whose direction can be controlled with an electric field.\cite{Fenni2008PRL} Benedek and Fennie showed that the ferroelectricity in FeTiO$_3$ is mostly attributable to displacement of $A$-site cations, with a minor contribution from off-center $B$-site cations.\cite{Fennie-JPCB2013} Varga \emph{et al}. synthesized $R3c$ FeTiO$_3$ at high pressure and confirmed the coexistence of weak ferromagnetism and ferroelectricity (multiferroicity) in this material.\cite{Varga2009}

More recently, Wang \emph{et al}.  predicted ZnFe$_{0.5}$Os$_{0.5}$O$_3$ to be a polar LiNbO$_3$-type perovskite with strong ferroelectricity ($\sim$ 54.7 $\mu$C/cm$^2$) caused by $A$-site displacements, at the same time showing that the ferroelectric ordering may be strongly coupled with ferrimagnetism above room temperature.\cite{Wang2015PRL} Ye and Vanderbilt also identified ZnFe$_{0.5}$Os$_{0.5}$O$_3$ and other LiNbO$_3$-type perovskites such as LiZr$_{0.5}$Te$_{0.5}$O$_3$ as ferroelectrics by calculating energy profiles
of the ferroelectric reversal paths from first principles.\cite{Ye2016}

As the above outline suggests, considerable progress has been made in our understanding of the ferroelectric behavior of LiNbO$_3$-type oxide perovskites. Relatively little attention has been paid, however, to LiNbO$_3$-type perovskites in which O is completely replaced with other 2nd row elements such as N or F.\cite{SmithCsPbF3-2015}

Here we report on the ferroelectric behavior of nitride perovskite LaWN$_3$, which was recently identified as a stable semiconductor based on high-throughput first-principles calculations.\cite{Rafael2015,Korbel2016}
These calculations predicted the ground state of LaWN$_3$ to have a LiNbO$_3$-type structure with $R3c$ symmetry, which can be derived from the aristotype (cubic) perovskite structure by two different phase transition sequences, viz., $Pm\bar{3}m$$\rightarrow$$R\bar{3}c$$\rightarrow$$R3c$ and $Pm\bar{3}m$$\rightarrow$$R3m$$\rightarrow$$R3c$.
The corresponding structural distortions associated with each phase transition can be interpreted in terms of symmetry-adapted modes.
In this study we use first-principles calculations to show that $R3c$ LaWN$_3$ exhibits robust ferroelectricity with a polarization of 61 $\mu$C/cm$^2$. Surprisingly, although the  Goldschmidt tolerance factor, $t$, is less than 1, the ferroelectric instability is found to be driven by $B$-site atom displacement rather than \emph{A}-site atom displacement, unlike the well-known ferroelectric LiNbO$_3$-type semiconductor ZnSnO$_3$\cite{Inaguma2008} and ferroelectric metal LiOsO$_3$.\cite{Xiang-PRB-2014,Liu2015}

The rest of the paper is organized as follows: In Sec. \ref{sec-II} we provide details of the first-principles calculation methods. The results are presented and discussed in Sec. \ref{sec-III}, and conclusions are summarized in Sec. \ref{sec-IV}.

\section{COMPUTATIONAL DETAILS}\label{sec-II}
First-principles density functional theory (DFT) calculations are performed using the Vienna \emph{ab initio} Simulation Package (VASP)\cite{Kresse1996,Kresse-PRB-1996} based on the projector-augmented wave method and a plane wave basis set.\cite{Blochl-PRB-1994} 5$p^6$5$d^1$6$s^2$ for La, 5$d^4$6$s^2$ for W, and 2$s^2$2$p^3$ for N are explicitly included as valence electrons in the pseudopotentials. The local density approximation (LDA) is used for exchange-correlation terms. A planewave cutoff of 520 eV is used in all cases. 13$\times$13$\times$13 and 7$\times$7$\times$7 Monkhorst-Pack \emph{k}-meshes\cite{Monkhorst1976} are used for unit cell and 2$\times$2$\times$2 supercell calculations, respectively. All structures are fully relaxed until Hellmann-Feynman forces on each atom converge to below 10$^{-3}$ eV$\cdot$\AA$^{-1}$.

Phonon calculations using 2$\times$2$\times$2 supercells are carried out in which force constants are calculated based on density functional perturbation theory\cite{PRB-Lee-1997DFPT} as implemented in VASP. Phonon band structures and phonon densities of states at arbitrary $q$-vectors are computed using the code phonopy.\cite{Togo2015} In addition, except for the phases of LaWN$_3$ showing metallic properties, longitudinal-optical-transverse-optical splitting is included using a non-analytical term correction.\cite{Wang2010jpcm} Ferroelectric polarization is calculated using the Berry phase method.\cite{BP-Vanderbilt-PRB-1993} The effects of hybridization of electronic states on ferroelectric properties are examined by the orbital selective external potential (OSEP) method.\cite{WAN-OSEP,FANG2015}

Because of the well-known underestimation of the bandgap using standard DFT, in addition to LDA calculations we also use the hybrid exchange-correlation functional of Heyd, Scuseria, and Ernzerhof (HSE) \cite{Heyd2003} in a few instances. In these cases, Brillouin zone sampling of the 10-atom unit cell is limited to a 5$\times$5$\times$5 \emph{k}-mesh grid.

\section{RESULTS AND DISCUSSION} \label{sec-III}
\subsection{Lattice dynamics of the cubic phase}
The Goldschmidt tolerance factor for perovskite compounds ${ABX}$$_3$ is defined as $t$ = $\frac{R_{\rm A} + R_{\rm X}}{\sqrt{2}(R_{\rm B} + R_{\rm X})}$, where $R_{\rm A}$, $R_{\rm B}$, and $R_{\rm X}$ are the radii of ions on $A$, $B$, and $X$ sites, respectively.
The tolerance factor gives an indication of the stability and amount of distortion of the structure\cite{tolerance1-natcomm2014,PRB-Rappe-2006} relative to the cubic aristotype, for which $t$ = 1. If $t$ $\neq$ 1, octahedral rotations or off-center displacements are expected to occur to optimize the coordination environments of $A$-site and $B$-site atoms. In the case of LaWN$_3$, the ionic radii of La and W in 12- and 6-fold coordination ($R_{\rm La}$ = 1.36 \AA{} and  $R_{\rm W}$ = 0.6 \AA), respectively and the ionic radius of N ($R_{\rm N}$ = 1.46 \AA{} for 4-fold coordination) give $t$ = 0.969; the cubic form of LaWN$_3$ is thus not expected to be the ground-state structure, as the large deviation from $t$ = 1 indicates distortions to low-symmetry structures should be more energetically favorable.

\begin{figure}
  \centering
  \includegraphics[width=0.75\linewidth]{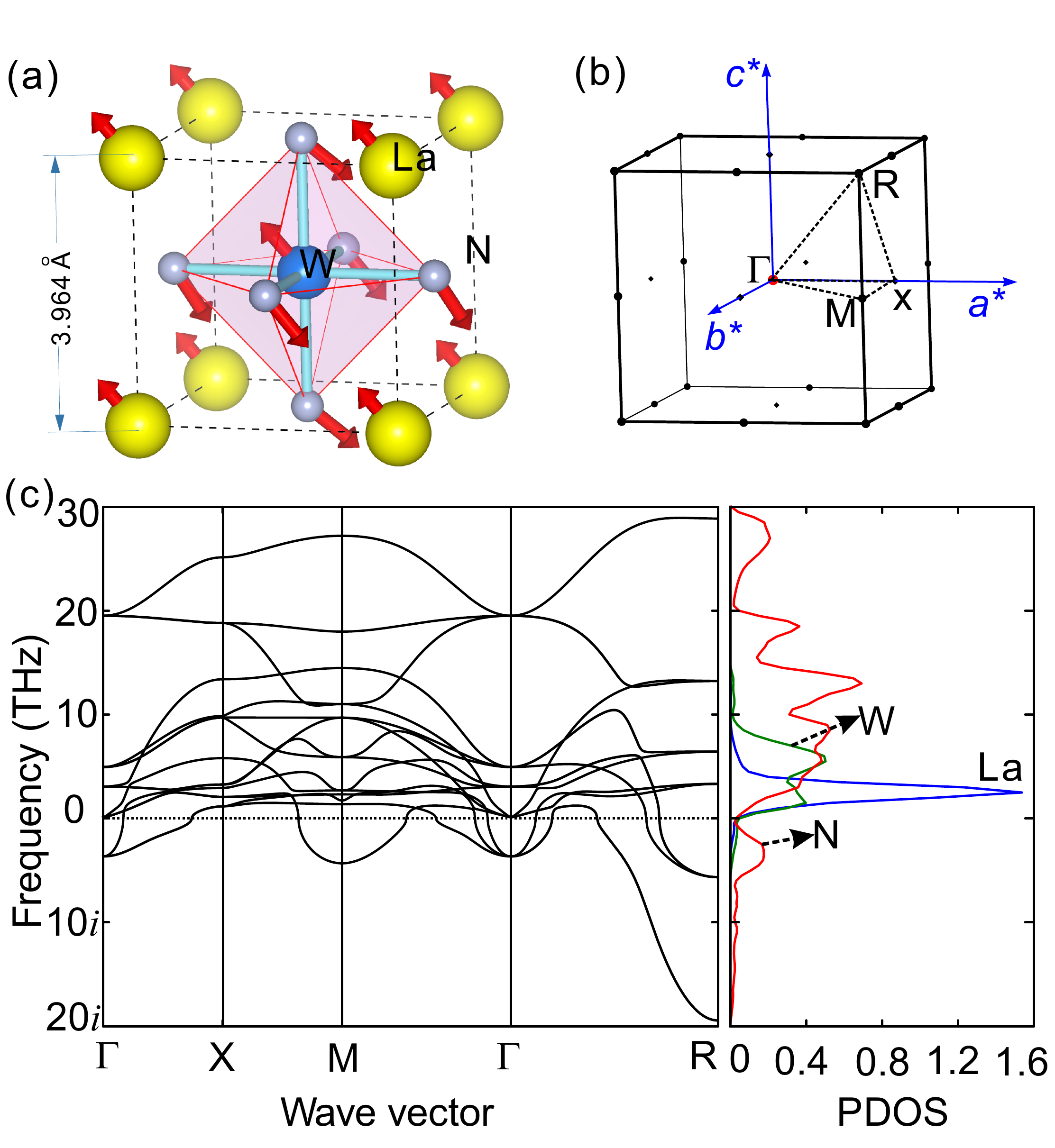}\vspace{-2pt}
  \caption{(Color online) (a) The unit cell of cubic LaWN$_3$ (space group $Pm\bar{3}m$); (b) the Brillouin zone of a face-centred cubic lattice showing the high-symmetry $\Gamma$-$X$-$M$-$\Gamma$-$R$ path used in phonon calculations. $a^*$, $b^*$, and $c^*$ are the primitive reciprocal
lattice vectors; and (c) phonon dispersion curves along the high-symmetry $k$-path and the projected phonon densities of states (PDOSs) normalized to the primitive cell. In the PDOS plot, blue, green, and red curves represent contributions of La, W, and N atoms, respectively.}\label{pm-3m} 
\end{figure}

DFT calculations to assess the structural stability of cubic LaWN$_3$ (space group $Pm\bar{3}m$) were performed using the LDA functional.
The unit cell of cubic LaWN$_3$ is illustrated in Fig. \ref{pm-3m}(a), and the Brillouin zone path used in phonon dispersion calculations is shown in Fig. \ref{pm-3m}(b).
Structural relaxation resulted in a lattice constant of 3.964 \AA.

Phonon dispersion curves calculated along high-symmetry lines $\Gamma$-$X$-$M$-$\Gamma$-$R$ in the Brillouin zone, and partial phonon density of states for each element, are shown in Fig. \ref{pm-3m}(c). The imaginary vibration modes around $\Gamma$, $M$, and $R$ points indicate that cubic LaWN$_3$ is dynamically unstable. Strongly unstable antiferrodistortive modes at $R$ (20.78$i$ THz) and $M$ (4.61$i$ THz) points correspond to rotation of WN$_6$ octahedra, with the instability region in the projected phonon density of states (PDOS) plot almost exclusively associated with N motions.
Three transverse optical modes have large imaginary frequencies (3.879$i$ THz) at the $\Gamma$ point, suggesting the possibility of a ferroelectric transition because the eigenvectors of the $\Gamma$ point modes correspond to a displacement pattern in which N anions move in the opposite direction to the cations. The directions of the displacement modes are indicated by red arrows in Fig. \ref{pm-3m}(a).

\subsection{Prediction of low-symmetry phases}
Both the high instability of the cubic phase and Goldschmidt's rule point to LaWN$_3$ undergoing one or more structural phase transitions to lower-symmetry structures. By freezing in the unstable modes (i.e., the ferroelectric mode at $\Gamma$ and antiferrodistortive modes at $R$ and $M$) singly and in combination, we examined several low-symmetry phases, indentifying the $R3c$ phase as having the lowest lattice energy after structure relaxation.

The energy and structural parameters of the low-symmetry phases are reported in Table \ref{tab:searchphase}, with data of cubic LaWN$_3$ included for comparison. Based on the calculated lattice energies, the ground-state structure most likely to have $R3c$ symmetry, and other structures should be dynamically unstable at low temperatures. LaWN$_3$ with orthorhombic $Pnma$ symmetry was also calculated as this is the most commonly observed space group for oxide perovskites\cite{Fennie-JPCB2013}. In these cases, under-coordination of the $A$-site cations associated with low tolerance factors leads to octahedral rotations and antipolar $A$-site displacements which can ultimately suppress the ferroelectricity\cite{Benedek2012JSSC} The results in Table \ref{tab:searchphase} show, however, that in the case of LaWN$_3$, the $Pnma$ structure is higher in energy than the $R3c$ structure, resulting in the survival of ferroelectricity in LaWN$_3$ with the $R3c$ form.

Furthermore, in the calculations we also find strongly unstable modes at the $\Gamma$ point  in the case of the $Pnma$ structure (see Fig. S1(a) in Supplementary Material). By freezing in the unstable mode at the $\Gamma$ point, the structure transforms into a $Pna2_1$ structure. As seen in Table \ref{tab:searchphase}, the energy of the $Pna2_1$ structure is only 28.12 meV higher than that of $R3c$ structure, indicating $Pna2_1$ LaWN$_3$ is an alternative metastable phase. As given by the phonon dispersion curves of $Pna2_1$ LaWN$_3$ (see Fig. S1(b) in Supplementary Material), the imaginary frequencies are disappeared. Since $R3c$ LaWN$_3$ is the ground-state structure, we focus on the symmetry-related transitions between the three rhombohedral phases to examine the ferroelectric behavior of LaWN$_3$ in detail. Details of the structural transitions and electronic properties of $Pna2_1$ LaWN$_3$ are provided as Supplementary Material.

\begin{figure}
  \centering
  \includegraphics[width=0.75\linewidth]{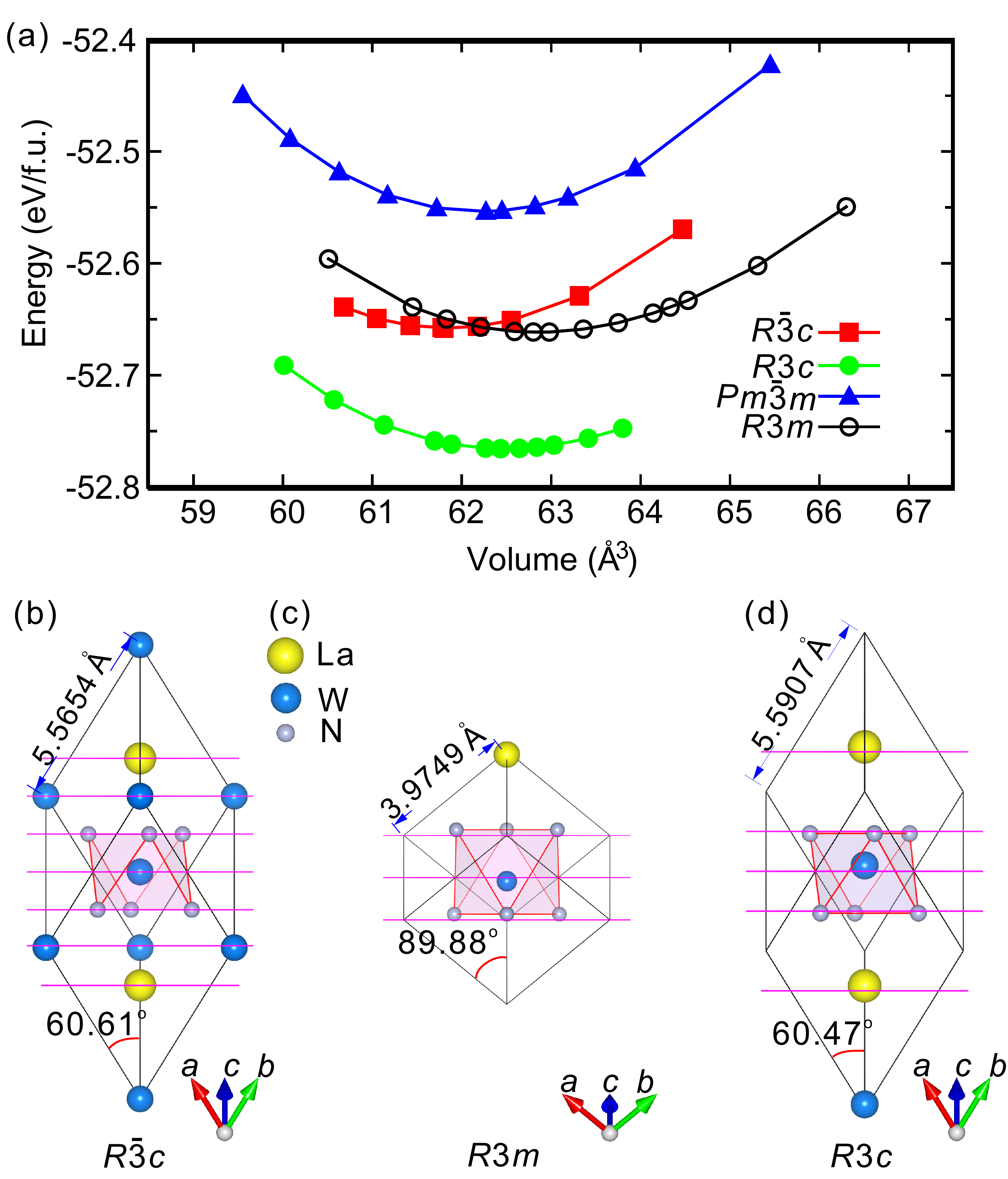}\vspace{-2pt}
  \caption{(Color online) (a) Total energy per formula unit as a function of volume per formula unit for $Pm\bar{3}m$, $R\bar{3}c$, $R3m$, and $R3c$ phases of LaWN$_3$. (b) The 10-atom unit cell of the $R\bar{3}c$ structure. (c) The 5-atom unit cell of the $R3m$ structure. (d) The 10-atom unit cell of the ground-state structure ($R3c$). Magenta lines indicate (111) planes. The projection vector for each structure is the [111] direction of the pseudocubic lattice. In the $R\bar{3}c$ structure, W is located at the inversion center, while $R3m$ and $R3c$ structures are non-centrosymmetric, with the W atom displaced from the inversion center in each case. }\label{VE} 
\end{figure}

\begin{table*}[!ht]
\centering
\caption{Structure parameters, Wyckoff positions, and relative stabilities of different phases of LaWN$_3$ from DFT-LDA calculations. Each phase is identified by its space group. Stabilities are reported as differences in 
	total energy relative to the $Pm\bar{3}m$ phase. Note that the published version at Phys. Rev. B 95, 014111 (2017) has a typo error in the Wyckoff coordinates of W, which has been fixed here.}
\begin{center} \footnotesize
\begin{tabular}{l  c c c c c c c c c c c l|c|c|c|c|} \hline \hline %
Phase   &   & Structure parameters & & & Atom & Wyckoff & & Coordinates & & Energy  \\
     &               $a$ & $b$ & $c$ & $\alpha$ & & site & $x$ & $y$ &$z$  & $\Delta$E\\
     &               (\AA) & (\AA)  & (\AA) &  ($^\circ$)    & & & & & & (meV/f.u.)      \\
		\hline
$Pm\bar{3}m$    & $3.9635$ & $-$ & $-$ & $90$ & La & $1a$ & $0.0$ & $0.0$ & $0.0$ & $0.0$\\ 
& & & & & W  & $1b$ & $0.5$ & $0.5$ & $0.5$ \\
& & & & &  N & $3c$ & $0.0$ & $0.5$ & $0.5$ \\
$P4/mbm$  &   $5.5590$ & $-$ & $4.0023$ & $90$ & La & $2d$ & $0.0$ & $0.5$ & $0.0$ & $-53.35$\\ 
& & &  & & W  & $2b$ & $0.0$ & $0.0$ & $0.5$ \\
& & & &  & N1 & $4h$ & $0.28917$ & $0.78917$ & $0.5$ \\ 	
& & & & &  N2 & $2a$ & $0.0$ & $0.0$ & $0.0$ \\
$I4/mmm$  &  $5.6030$ & $-$ & $7.9231$ & $90$ &  La & $4d$ & $0.0$ & $0.5$ & $0.25$ & $-0.51$    \\ 
& & &  & & W1  & $2b$ & $0.0$ & $0.0$ & $0.5$ \\
& & &  & & W2  & $2a$ & $0.0$ & $0.0$ & $0.0$ \\
& & & &  & N1 & $8h$ & $0.24868$ & $0.24868$ & $0.0$ \\ 	
& & & &  & N2 & $4e$ & $0.0$ & $0.0$ & $0.25263$ \\ 
$I4/mcm$  &  $5.6029$ & $-$ & $7.9238$ & $90$ & La & $4a$ & $0.0$ & $0.0$ & $0.25$ & $-0.25$ \\ 
& & &  & & W  & $4d$ & $0.0$ & $0.5$ & $0.0$ \\
& & & &  & N1 & $8h$ & $0.25171$ & $0.75171$ & $0.0$ \\
& & & &  & N2 & $4b$ & $0.0$ & $0.5$ & $0.25$ \\
$Pnma$  &  $5.6010$ & $7.9142$ & $5.6386$ & $90$ & La & $4c$ & $-0.01466$ & $0.25$ & $0.49822$ & $-105.93$ \\ 
& & &  & & W  & $4a$ & $0.0$ & $0.0$ & $0.0$ \\
& & & &  & N1 & $8d$ & $0.73057$ & $0.4712$ & $0.2328$ \\
& & & &  & N2 & $4c$ & $0.00373$ & $0.25$ & $-0.05553$ \\
$Pna2_1$  &  $5.5703$ & $ 5.5992$ & $8.0137$ & $90$ & La & $4a$ & $0.48115$ & $0.49701$ & $0.75$ & $-184.10$ \\ 
& & &  & & W  &  $4a$  &  $0.51666$ & $0.00417$ & $0.00903$   \\
& & & &  & N1 & $4a$ &  $0.23257$ & $0.23004$ & $0.51592$ \\
& & & &  & N2 & $4a$ & $0.71842$ & $0.27548$ & $-0.04178$ \\
& & & &  & N3 & $4a$ & $0.50416$ & $-0.05530$ & $0.73893$ \\
$R3m$     &  $3.9749$ & $-$ & $-$  & $89.8841$ & La & $1a$ & $-0.01358$ & $-0.01358$ & $-0.01358$ &  $-107.87$    \\ 
& & &  & &W  & $1a$ & $0.4863$ & $0.4863$ & $0.4863$ \\
& & & &  &N & $3b$ & $0.52566$ & $0.52566$ & $0.01593$ \\
$R3c$   &  $5.5907$ & $-$ & $-$ & $60.4666$  & La & $2a$ & $0.26039$ & $0.26039$ & $0.26039$ & $-212.22$  \\ 
& & &  & &W  & $2a$ & $0.01239$ & $0.01239$ & $0.01239$ \\
& & & &  &N & $6b$ & $0.69773$ & $0.79658$ & $0.24290$ \\
$R\bar{3}c$   &  $ 5.5654$ & $-$ & $-$ & $60.6077$ & La & $2a$ & $0.25$ &  $0.25$ &$0.25$ &  $-103.94$  \\ 
& & &  & &W  & $2b$ & $0.0$ & $0.0$ & $0.0$ \\
& & & &  &N & $6e$ & $0.70144$ & $0.79856$ & $0.25$ \\
\hline
\hline
\end{tabular}
\end{center}
\label{tab:searchphase}
\end{table*}
The relative stabilities of the $R\bar{3}c$, $R3m$, and $R3c$ phases were examined by plotting their energies per formula unit (f.u.) as a function of volume per f.u., as shown in Fig. \ref{VE}(a). The lattice parameter corresponding to the minimum energy for each low-symmetry phase matches the corresponding lattice constant in Table \ref{tab:searchphase} obtained by freezing in the unstable modes to within about $\pm$0.1$\%$. Therefore, both Fig. \ref{VE}(a) and Table \ref{tab:searchphase} unambiguously prove that $R3c$ is the ground-state structure. 

The tilting systems in the different phases of LaWN$_3$ can be described using Glazer's notation.\cite{Glazer1972} In rhombohedral nitride perovskites ${AB}$N$_3$, each $B$N$_6$ octahedron is tilted about its triad axis, and this distortion can be described as $a^{-}$$a^{-}$$a^{-}$ tilting since adjacent octahedra rotate in opposite directions. In the case of LaWN$_3$, both $R3c$ and $R\bar{3}c$ structures belong to this tilting system. The tilting originates from a delicate balance between the rigidity of the WN$_6$ octahedra and large size difference between La and W ions, and the zone center soft mode. The $R3c$ phase is ferroelectric because of the non-centrosymmetry of its lattice; conversely the $R\bar{3}c$ phase remains non-polar, with $B$ cations located on inversion centers. The $R\bar{3}c$ structure is derived from the $Pm\bar{3}m$ structure purely by an $a^{-}$$a^{-}$$a^{-}$ tilting pattern, while the $R3c$ structure is a combination of $a^{-}$$a^{-}$$a^{-}$ tilting and displacements of ions along pseudocubic $<$111$>$ directions. Unlike the $R3c$ and $R\bar{3}c$ phases, octahedral tilting does not occur in $R3m$ LaWN$_3$. However, the space inversion symmetry of the $R3m$ structure is broken, just as it is in the $R3c$ structure, resulting in a net polarization in the \emph{z} direction.

The calculated rhombohedral structures of the $R\bar{3}c$, $R3m$, and $R3c$ phases are illustrated in Figs. \ref{VE}(b), (c), and (d), respectively. Although no net dipole moment can be induced in the case of the $R\bar{3}c$ phase, the non-centrosymmetry of the $R3c$ and $R3m$ structures makes them candidates for exhibiting ferroelectric behavior.
Comparison of the lattice energies in Table \ref{tab:searchphase} and Fig. \ref{VE} shows that the energy of the $R3m$ phase is only 3.9 meV per f.u. lower than that of the $R\bar{3}c$ phase. This small energy difference suggests that the two phases will coexist at low temperatures, not unlike the situation for AgNbO$_3$.\cite{Moriwake2016}

\begin{figure}
  \centering
  \includegraphics[width=0.75\linewidth]{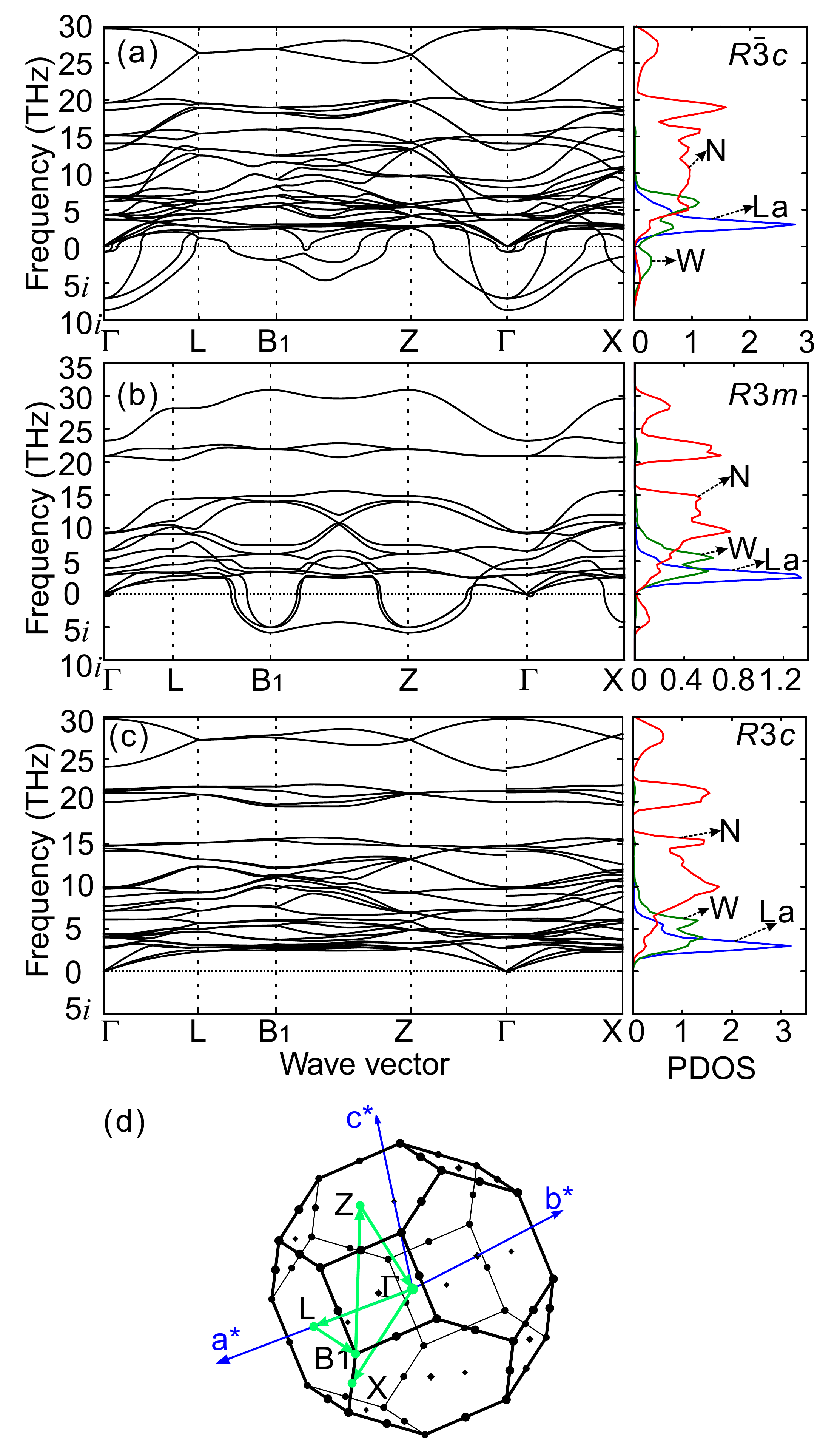}\vspace{-2pt}
  \caption{(Color online) Phonon dispersion curves and projected phonon densities of states (PDOSs) normalized to their respective primitive cells for (a) $R\bar{3}c$, (b) $R3m$, and (c) $R3c$ structures; and (d) the Brillouin zone of the rhombohedral lattice. Blue, green, and red curves in the PDOS plots represent the contributions of La, W, and N atoms, respectively. In (d), the dispersion path (green) passes through special high-symmetry points $\Gamma$, $L$, $B_1$, $Z$, and $X$. Reciprocal lattice vectors are shown as blue arrows.}
  \label{rho-phonon} 
\end{figure}

\parskip=2pt The relative stabilities of the three rhombohedral phases can be understood in terms of their vibrational mode frequencies. Figures. \ref{rho-phonon}(a), (b), and (c) show the calculated phonon dispersion curves and projected phonon densities of states for each element for $R\bar{3}c$, $R3m$, and $R3c$ structures, respectively.
The high symmetry path of the first Brillouin zone used in the calculations is shown in Fig. \ref{rho-phonon}(d). The $R3c$ structure does not exhibit any unstable modes, while the $R\bar{3}c$ and $R3m$ structures exhibit strongly unstable modes, confirming that the $R3c$ structure is the ground state. Although the projected PDOSs of the $R3m$ phase are similar to those of the $R\bar{3}c$ phase, the unstable modes are different. Specifically, the imaginary frequencies in the case of the $R3m$ structure occur at points $Z$ and $B_1$, corresponding to displacements of N atoms, whereas the centrosymmetric $R\bar{3}c$ phase exhibits an unstable mode at the $\Gamma$ point that is associated with displacements of all three types of atoms.

$R3c$ LaWN$_3$ has not yet been synthesized, so to assess its chemical stability we calculated the formation enthalpy relative to component phases LaN and WN$_2$, whose structures were taken from Ref. \citen{Khitrova1961} and Ref. \citen{Olcese1979} respectively, according to $\Delta$$H^f$(LaWN$_3$) = $E$(LaWN$_3$) - $E$(LaN) - $E$(WN$_2$), where $E$(LaWN$_3$), $E$(LaN), and $E$(WN$_2$) are the total energies per f.u. of LaWN$_3$, LaN, and WN$_2$, respectively. The calculated formation enthalpy of -11.617 eV per f.u. suggests that it should be possible to synthesize LaWN$_3$ in its $R3c$ form.
\subsection{Structural phase transitions}
The structural phase transition pathways and mechanisms between the three rhombohedral phases were also investigated using first-principles calculations.

Two possible transition sequences with increasing temperature from the rhombohedral phases to the $Pm\bar{3}m$ phase were identified from the symmetry relations between the different phases, viz. from the polar $R3c$ phase with a$^-$a$^-$a$^-$ tilting to the $Pm\bar{3}m$ phase via either the $R\bar{3}c$ or $R3m$ structure. Similarly, upon cooling, the phase transition sequence could be either $Pm\bar{3}m$$\rightarrow$$R\bar{3}c$$\rightarrow$$R3c$ or $Pm\bar{3}m$$\rightarrow$$R3m$$\rightarrow$$R3c$.
To examine the transition mechanisms in more detail, we calculated the total energy of each phase as a function of the mode amplitude for each of the possible pathways. Here the amplitude $Q$ of a specified  mode is defined as $Q$ = $\sqrt{\sum\limits_{i} \zeta{_{i\alpha}^2}}$, where $\zeta_{i}$ denotes the displacement of atom $i$ in the direction $\alpha$ from its equilibrium position in the parent structure.

\begin{figure}
  \centering
  \includegraphics[width=0.75\linewidth]{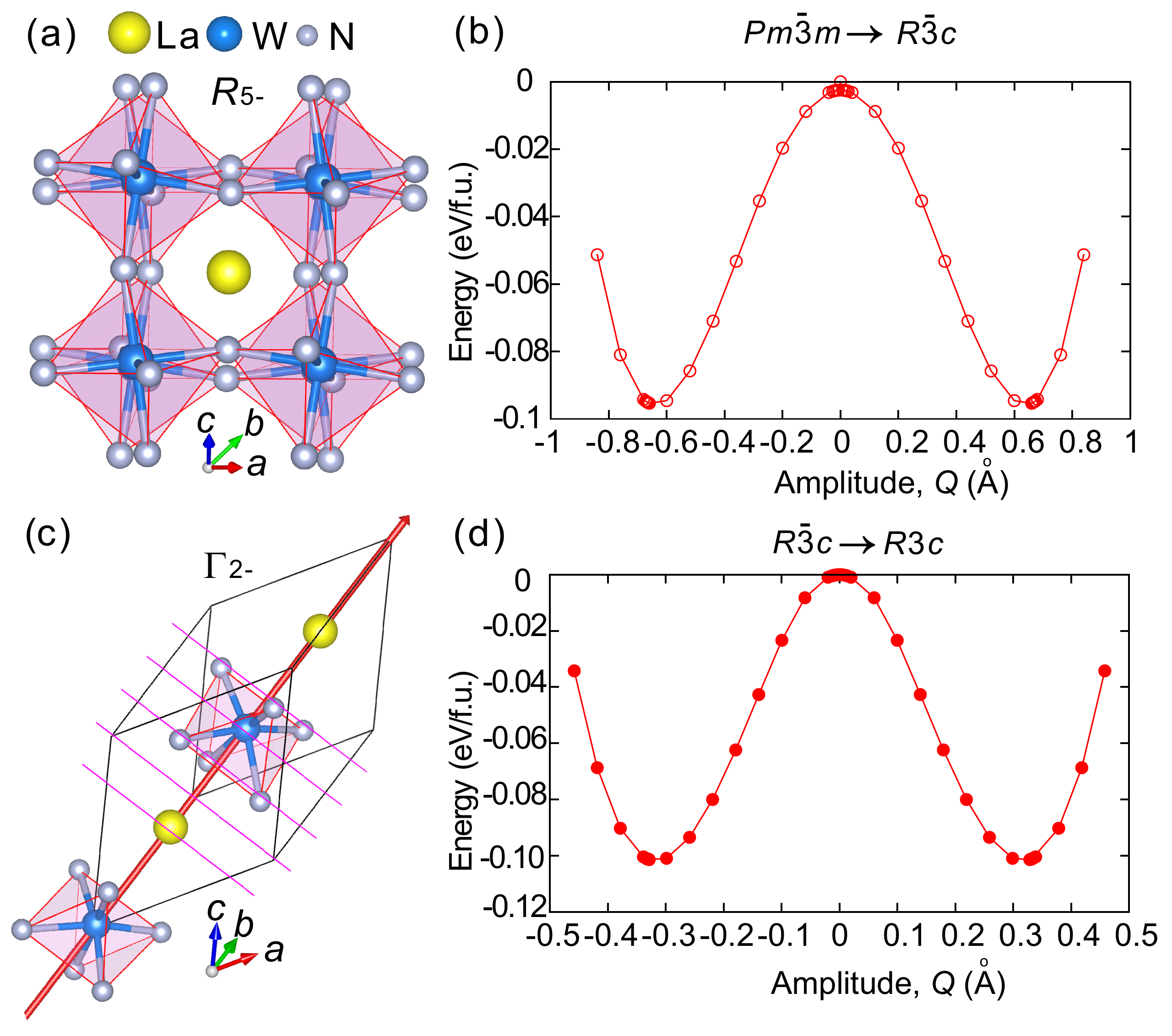}\\
  \caption{(Color online) (a) Schematic of the reference structure with only the $R_{5-}$ mode frozen. (b) Energy-amplitude curve for the $Pm\bar{3}m$$\rightarrow$$R\bar{3}c$ transition, with energies taken relative to that of the $Pm\bar{3}m$ structure. (c) Schematic representation of the reference structure with $\Gamma_{2-}${} mode frozen. (d) Energy-amplitude curve for the $R\bar{3}c$$\rightarrow$$R3c$ transition, with energies relative to that of the $R\bar{3}c$ structure.}
  \label{transition1} %
\end{figure}

\emph{Pathway $Pm\bar{3}m$$\rightarrow$$R\bar{3}c$$\rightarrow$$R3c$}. The transition pathway $Pm\bar{3}m$$\rightarrow$$R\bar{3}c$ is associated with the $a^{-}$$a^{-}$$a^{-}$ tilting mode with irreducible representation $R_{\rm 5-}$. The tilting mode caused by the rotation of a WN$_6$  octahedron is illustrated in Fig. \ref{transition1}(a). The amplitude of the distortion is around 0.67 \AA. Figure \ref{transition1}(b) shows the associated energy-amplitude curve. The energy to stabilize the WN$_6$ octahedron is about 95.36 meV per f.u., indicating that the transition from the $R\bar{3}c$ phase to the cubic $Pm\bar{3}m$ phase occurs at a temperature of 1106 K. The transition from $R\bar{3}c$ to $R3c$ is related to the irreducible representation $\Gamma_{2-}${} with amplitude 0.33 \AA. The $\Gamma_{2-}${} mode involves displacement of W and La atoms in the pseudocubic [111] direction and bending of N-W-N bonds, as illustrated in Fig. \ref{transition1}(c). The double-well shape of the energy-amplitude curve in Fig. (d) indicates the $R\bar{3}c$$\rightarrow$$R3c$ ferroelectric transition involves spontaneous symmetry breaking. The energy associated with the $\Gamma_{2-}${} mode is about 101.4 meV per f.u., corresponding to a temperature increase of 1177 K.

\begin{figure}
  \centering
  \includegraphics[width=0.75\linewidth]{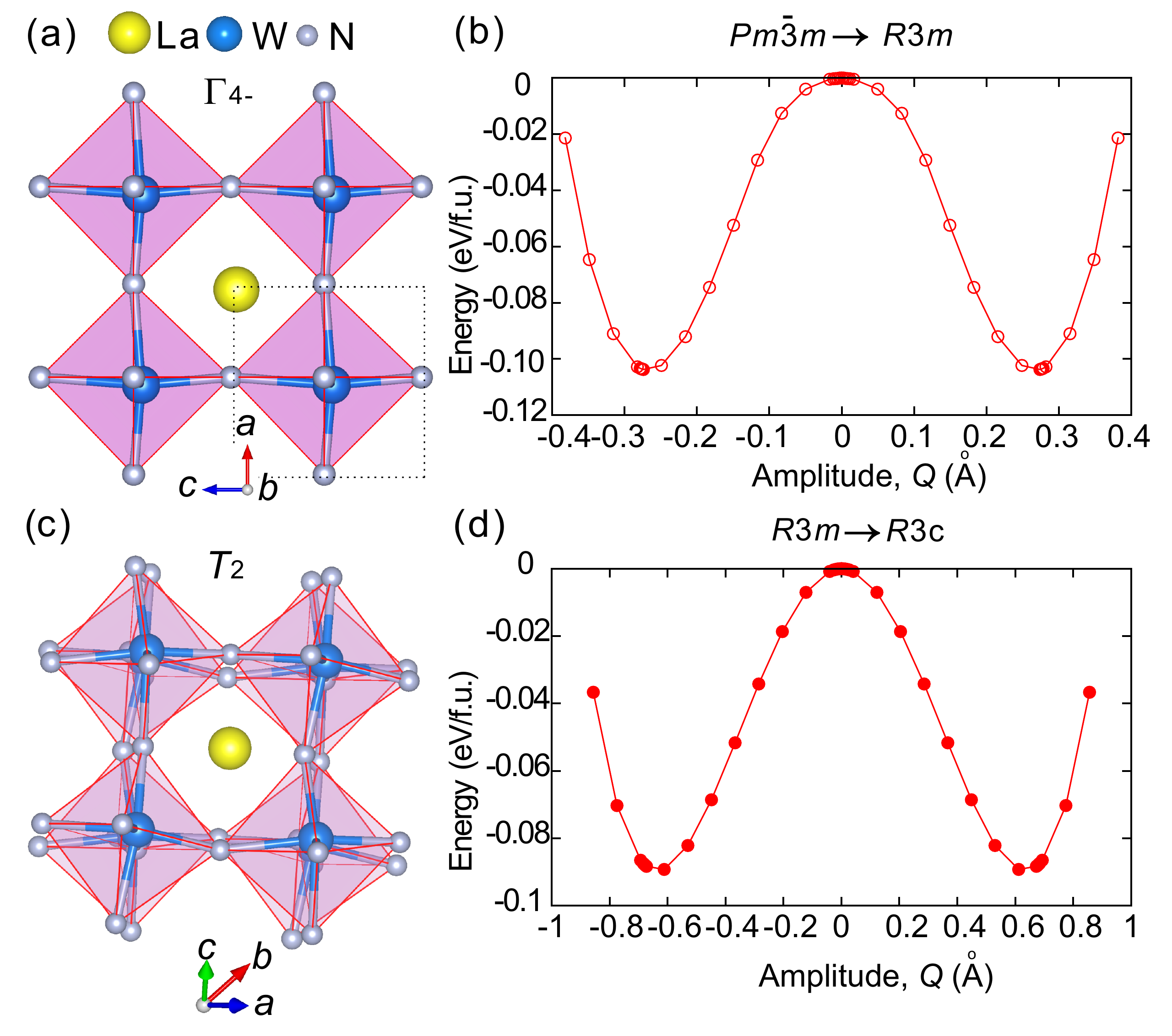}\\
  \caption{(Color online) (a) Schematic of the reference structure with frozen symmetry mode $\Gamma_{4-}$. (b) Energy-amplitude curve for the $Pm\bar{3}m$$\rightarrow$$R3m$ transition, with energies relative to that of the $Pm\bar{3}m$ structure. (c) Schematic of the reference structure with frozen $T_2$ mode. (d) Energy-amplitude curve for the $R3m$$\rightarrow$$R3c$ transition, with energies relative to that of the $R3m$ structure.}
  \label{transition2} %
\end{figure}

\emph{Pathway $Pm\bar{3}m$$\rightarrow$$R3m$$\rightarrow$$R3c$}. The symmetry-adapted modes associated with the $Pm\bar{3}m$$\rightarrow$$R3m$$\rightarrow$$R3c$ transition are shown in Fig. \ref{transition2}. A four-dimensional $\Gamma_{4-}${} distortion with an amplitude of 0.28 \AA{} results in a phase transition from the $Pm\bar{3}m$ structure to $R3m$ structure. The structure with the frozen $\Gamma_{4-}${} mode is shown in Fig. \ref{transition2}(a). This $\Gamma_{4-}${} mode is contributed by La-$T_{1u}$, W-$T_{1u}$, N-$A_{2u}$, and N-$E_u$. In the $R3m$ phase, no octahedral tilting occurs, and the inversion center vanishes on account of displacements of La, W, and N atoms along the pseudocubic [111] direction. The energy-amplitude curve in Fig. \ref{transition2}(b) indicates the phase transition from $R3m$ to $Pm\bar{3}m$ requires an energy of 103.75 meV per f.u.
Figures \ref{transition2}(c) and (d) show that the $R3c$ structure is stabilized by a $T_2$ distortion with a large amplitude of 0.68 \AA{}, resulting in an energy decrease of 89.24 meV per f.u. relative to $R3m$. The energy-amplitude curves show that the transition $R3m$$\rightarrow$$R3c$ is more energetically favorable than $R\bar{3}c$$\rightarrow$$R3c$, consistent with the results in Fig. \ref{VE}



\subsection{Dielectric and electronic properties}
The suitability of a ferroelectric for a particular application depends as much on its electronic properties as its ferroelectric behavior. Using the Berry phase method\cite{BP-Vanderbilt-PRB-1993}, we calculated the ground-state $R3c$ phase of LaWN$_3$ to have a spontaneous polarization of around 61 $\mu$C/cm$^2$. This polarization was calculated relative to an $R\bar{3}$c structure with the same lattice constants as the polar $R3c$ phase. The $R\bar{3}c$ reference phase was found to be metallic (no band gap), however, so we manually chose suitable switching paths to compute the allowed value of polarization, in the same manner as used by Neaton \emph{et al}. in their study of $R3c$ BiFeO$_3$.\cite{Neaton2005} Recently Sarmiento-P$\acute{\rm e}$rez \emph{et al}.\cite{Sarmiento-Perez2015} and  K{\"o}rbel \emph{et al}.\cite{Korbel2016} reported polarization values of 66 and 73 $\mu$C/cm$^2$, respectively, for LaWN$_3$, which are slightly larger than our value. This discrepancy can be attributed to the different lattice parameters used, since our calculated lattice energy and volume for the $R3c$ phase are 103.48 meV/f.u. and 3.315 \AA$^3$/f.u. smaller, respectively, than that of the structure used by Rafael \emph{et al}.\cite{Sarmiento-Perez2015} Also, K{\"o}rbel \emph{et al}. did not consider rotation of the WN$_6$ octahedra in their model,\cite{Korbel2016} which results in overestimation of the polarization.

\begin{figure}
  \centering
  \includegraphics[width=0.75\linewidth]{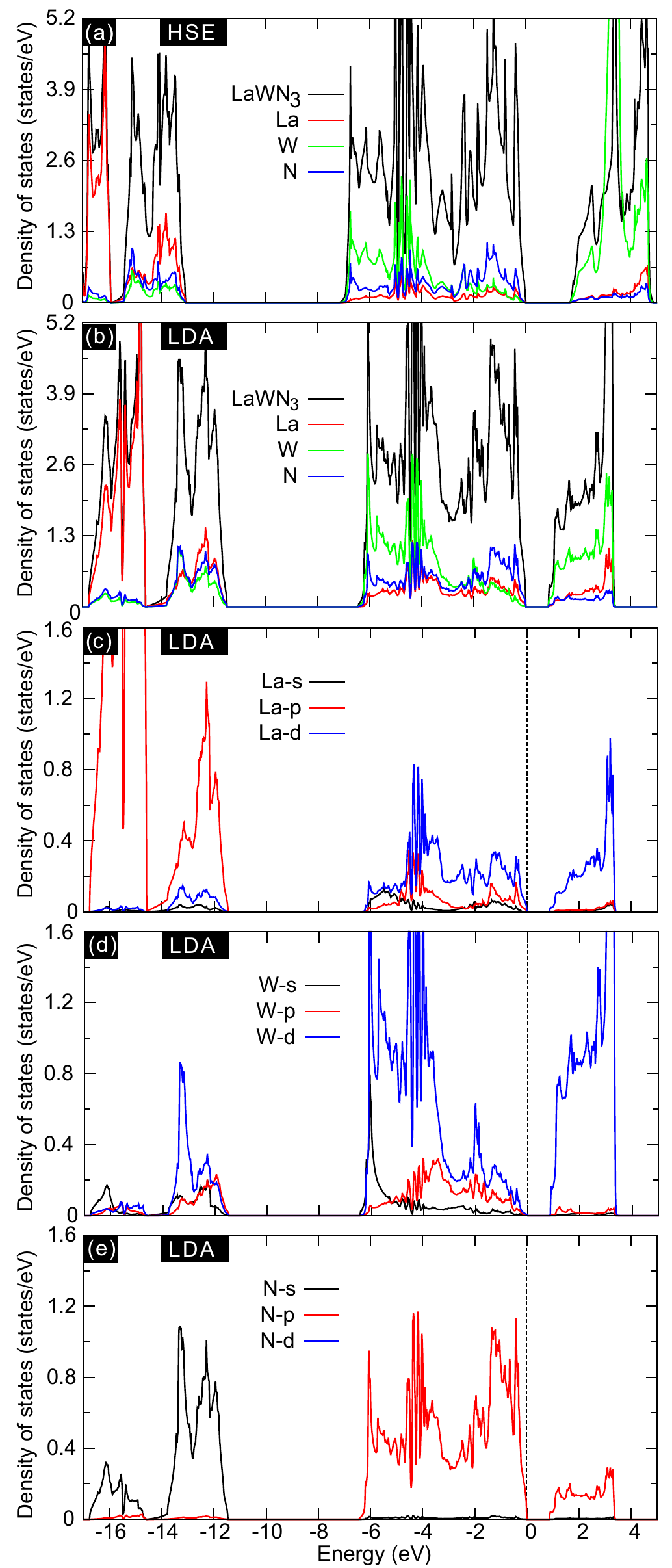}\\
  \caption{(Color online) Total and partial electronic densities of states of ferroelectric $R3c$ phase using the (a) HSE functional, and (b) LDA functional. The total densities of states are normalized to one formula unit. (c), (d), and (e) show the densities of states projected onto orbitals of La, W, and N atoms, respectively. The Fermi level is at 0 eV.}
  \label{GGA-HSE} %
\end{figure}

Figures \ref{GGA-HSE}(a) and (b) show the total and partial electronic densities of states of $R3c$ LaWN$_3$ calculated using HSE and LDA functionals, respectively. The corresponding band gaps are 1.72 and 0.81 eV. The HSE hybrid functional typically provides a more accurate estimate of the band gap in semiconductors,\cite{Heyd2005} but apart from the difference in gap width, the main features of the density of states according to the two methods are almost the same. This justifies our use of the computationally less-expensive LDA functional in calculations of phase stability and ferroelectric transitions.

Figures \ref{GGA-HSE}(c), (d) and (e) show the partial densities of states from the LDA calculations of La, W, and N atoms, respectively, split into their $s$, $p$, and $d$ component orbitals. The densities of states of La and W either side of the band gap mostly comprise their respective 5$d$ and 5$p$ states, and those of N are dominated by the 2$p$ states. The partial densities of states show La-5$d$, W-5$d$, W-5$p$, and N-2$p$ states undergo significant hybridization. Specifically, the energy separation of these electronic states determines the band gap. The strong cross-gap hybridization between N-2$p$ and unoccupied W-5$d$ states is essential for the onset of the ferroelectric instability, details of which are provided as Supplementary Material. In addition, the small contribution of La-6$s$ states to the bonding suggests that LaWN$_3$ is different to bismuth-based ferroelectric materials such as BiFeO$_3$ and BiMnO$_3$, in which the ferroelectric instability is caused by the stereochemical activity of the lone-pair electrons of the $A$-site Bi cations, which do not participate in chemical bonding.\cite{Seshadri2001,Khomskii2006}

\begin{figure}
  \centering
  \includegraphics[width=0.75\linewidth]{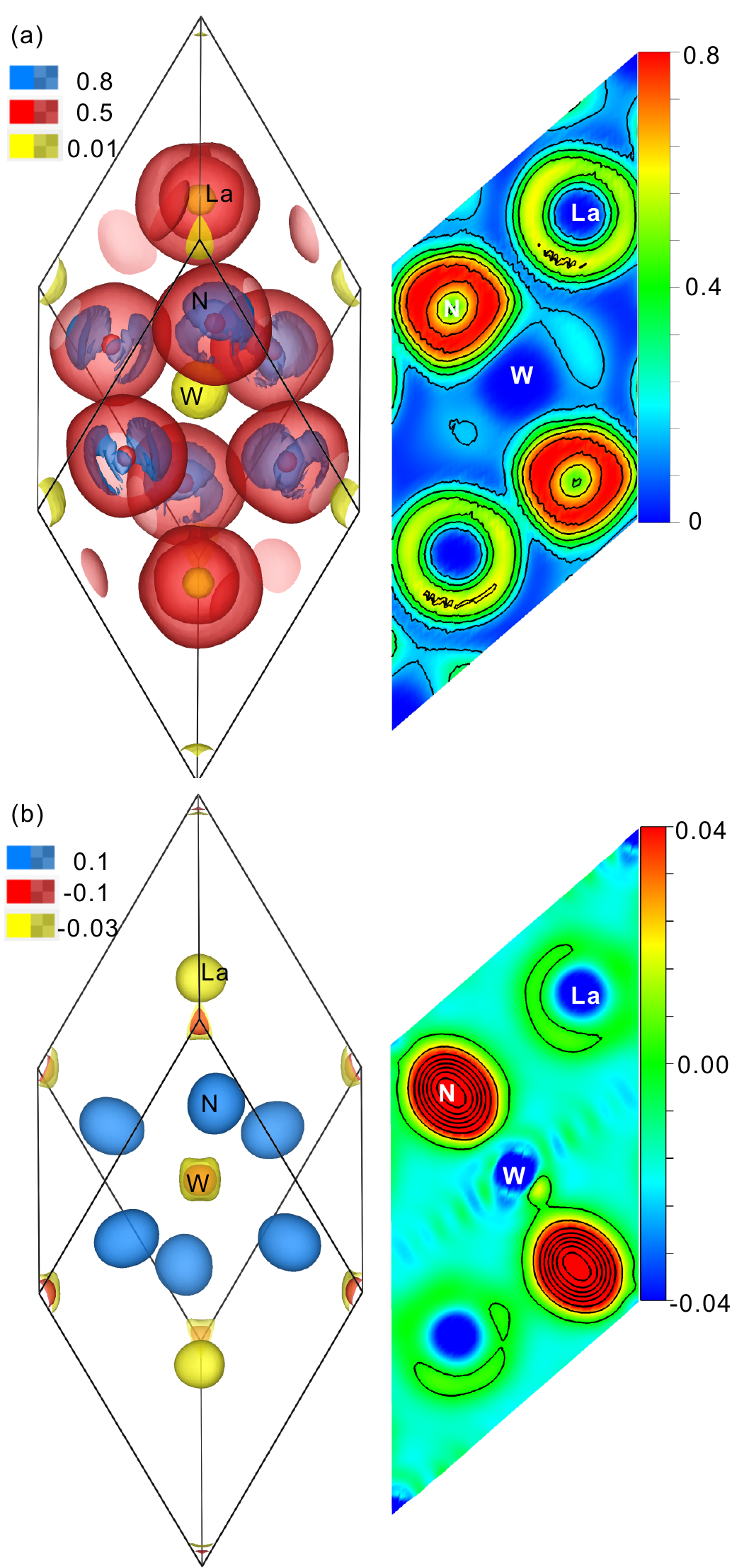}\\
  \caption{(Color online) (a) Maps of the electron localization function (ELF) in a unit cell of $R3c$ LaWN$_3$: left, 3D isosurface plot; right, 2D slice through the (1$\overline{\rm 1}$0) plane. Isosurface values are set at 0.8 (blue), 0.5 (red), and 0.01 (yellow). (b) Charge density difference maps in a unit cell of $R3c$ LaWN$_3$: left, 3D isosurface plot; right: slice through the (1$\overline{\rm 1}$0) plane. Isosurface values are set at 0.1 (blue), -0.1 (red) and -0.03 (yellow). Charge density differences were obtained by subtracting superposed atomic densities from the charge density of the $R3c$ LaWN$_3$ crystal.
}
  \label{ELF} %
\end{figure}

The electronic structure and bonding character of $R3c$ LaWN$_3$ can be appraised more quantitatively from plots of its electron localization function (ELF) shown in Fig. \ref{ELF}(a). The ELF provides an estimate of the probability of finding an electron at a given location.\cite{Becke1990} A value close to one means there is a high probability of an electron being found at that position, while a value of zero means electrons are fully delocalized or no electron is at that location; an ELF close to one-half corresponds to an electron-gas-like pair distribution. In Fig. \ref{ELF}(a), electron-gas-like regions (red) exist around La and N atoms. The charge distribution around La is also very uniform. This indicates that there are no localized lobe-shaped charge distributions (lone-pair electrons), in contrast to BiMnO$_3$, whose lone pairs are readily apparent in ELF plots.\cite{Seshadri2001}
In the case of LaWN$_3$, electrons are strongly localized around N, and the maximum value of the ELF between La and N is less than 0.2, suggesting that the La-N bond is mostly ionic. In contrast, electrons around W appear almost fully delocalized, so ELF values near W atoms are much smaller than near La atoms, indicating a higher covalency of the W-N bond. Based on the charge density difference maps in Fig. \ref{ELF}(b), more electrons are transferred from W to N than from La to N. Moreover, in the 2D slice in the right-hand panel, some electrons are localized between W and N atoms, which confirms W and N bond by sharing electron pairs.

Quantitative charge analyses were also carried out to evaluate the nature of the bonding, i.e., the amount of charge transfer between atoms, in $R3c$ LaWN$_3$. Born effective charges $Z^*$ were obtained using the density functional perturbation theory method. The diagonal entries in the tensor matrices for La, W, and N in Table \ref{BEC-table} are all much larger than their nominal ionic charges, consistent with LaWN$_3$ being strongly polarizable; in particular, the much larger Born effective charge of W suggests that W-N bonds are more covalent than La-N bonds. Similarly, Bader (static) charges calculated for La, W, and N
of +1.86e, +2.15e, and -1.34e, respectively, are smaller than their nominal charges by 38$\%$, 64$\%$, and 55$\%$, respectively, which also points to greater covalency in the W-N bonds, since the smaller the magnitude
of the Bader charge, the less ionized the atom is. The charge analyses thus confirm the trends observed from ELF analysis and charge density difference maps.
\begin{table*}[!ht]
\centering
\caption{Eigenvalues of calculated Born effective charge tensors $Z^*$ for La, W, and N in $R3c$ LaWN$_3$.}
\begin{center} \footnotesize
\begin{tabular}{l c c c c c c c  c c c c c c c l|c||c||c||c||c||c||c||c||c||c||c||c||c||c||c||c||c|} \hline \hline
Atom   &  & Coordinates  &  & & & & & Eigenvalues \\ 
     &    $x$ & $y$  & $z$ &  $Z^*_{xx}$ & $Z^*_{yy}$ & $Z^*_{zz}$ &$Z^*_{xy}$ &$Z^*_{xz}$ &$Z^*_{yx}$ &$Z^*_{xz}$ &$Z^*_{zx}$ &$Z^*_{zy}$    \\
		\hline
    La &  $0.260$ & $0.260$ & $0.260$ & $4.517$    &  $4.517$   &  $4.107$   & $0.436$   &  $0$  & $-0.437$   &   $0$  &  $0$   &  $0$ \\
    La & $0.760$ & $0.760$ & $0.760$ & $4.517$    &  $4.517$   &  $4.107$   & $-0.436$  & $0$ &   $0.437$  &    $0$ &  $0$   &  $0$ \\
    W & $0.012$ &  $0.012$ &  $0.012$ & $10.571$   & $10.571$   &  $6.844$   & $-0.775$  &   $0$ &  $0.774$   &   $0$  &  $0$   &  $0$ \\
    W & $0.512$ &  $0.512$ & $0.512$  & $10.571$   &  $10.571$  &  $6.844$   & $0.775$   &  $0$  & $-0.774$   &   $0$  &  $0$   &  $0$ \\
    N & $0.698$ &  $0.797$ & $0.243$  & $-3.204$   & $-6.856$   & $-3.650$   & $0.270$   &  $0.139$  & $0.566$    & $-1.882$   & $0.255$  & $-1.753$   \\
    N & $0.243$ &  $0.698$ & $0.797$  & $-5.581$   &  $-4.478$  & $-3.651$   & $-1.939$  &  $1.561$ &  $-1.642$  &    $1.061$ &  $1.391$   &  $1.098$   \\
    N & $0.797$ &  $0.243$ & $0.698$  & $-6.304$   &  $-3.754$  & $-3.651$   & $1.224$   & $-1.670$  &  $1.521$   &   $0.821$  &  $-1.646$  &   $0.655$  \\
    N & $0.297$ &  $0.198$ & $0.743$  & $-3.203$   &  $-6.856$  & $-3.651$   & $-0.270$  &  $-0.139$ &  $-0.566$  &   $-1.882$ & $ -0.255$  &  $-1.753$  \\
    N & $0.743$ &  $0.297$ & $0.198$  & $-6.304$   & $-3.754$   & $-3.651$   & $-1.225$  &   $1.670$ & $-1.521$   &   $0.821$  &  $1.646$   &  $0.656$   \\
    N & $0.198$ &  $0.743$ & $0.297$  & $-5.581$   & $-4.478$   & $-3.651$   & $1.939$   & $-1.561$  & $1.642$   &  $1.062$   &  $-1.391$  &   $1.098$  \\
\hline
\hline
\end{tabular}
\end{center}
\label{BEC-table}
\end{table*}

\begin{figure}
  \centering
  \includegraphics[width=0.75\linewidth]{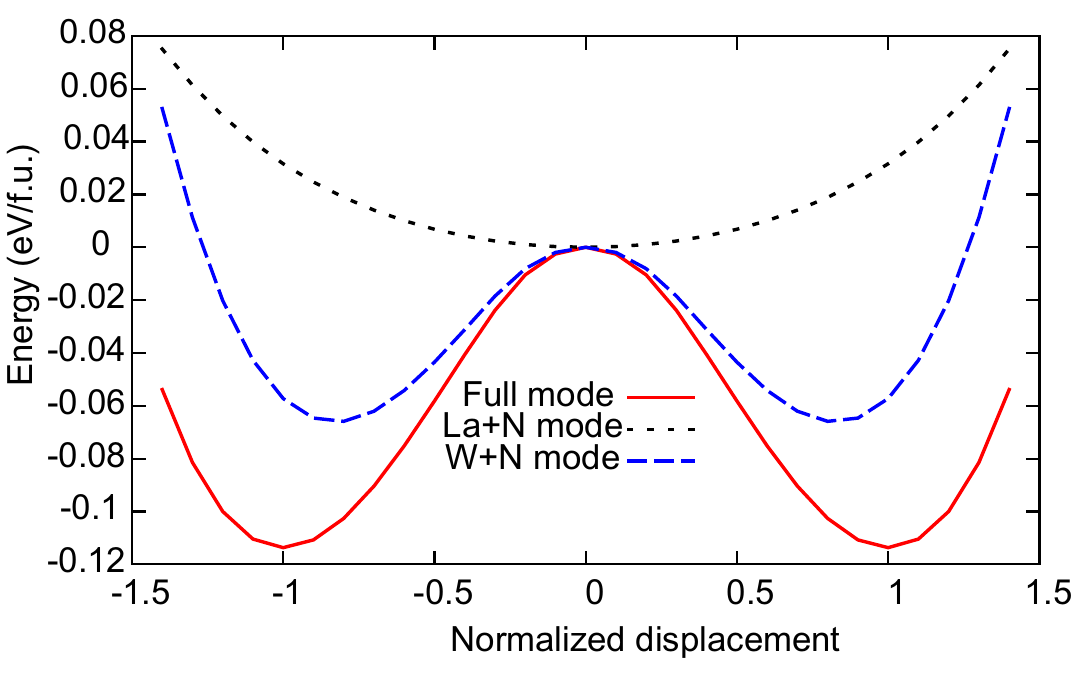}\\
  \caption{(Color online) Energy as a function of normalized displacement for full mode (red solid curve), La+N partial mode (black dot curve), and W+N partial mode (blue dash curve) in LaWN$_3$. The amplitude of the ion displacements along $<$111$>$ directions are normalized, i.e., 1 corresponds to the ferroelectric state with $R3c$ symmetry, and 0 corresponds to the paraelectric state with $R\bar{3}c$ symmetry.}
  \label{lambda} %
\end{figure}

To probe the microscopic mechanism of the ferroelectric phase transition in LaWN$_3$ further, we calculated the potential energy profile during polarization switching. A combination of soft modes at the $\Gamma$ point (i.e., $\Gamma_{2-}${} and $\Gamma_{1+}${} modes) was used as the full mode to which the displacements of all atoms were compared. We also examined the La+N mode, in which atomic displacements of W are neglected, to study the partial contribution of La ions; similarly the W+N mode was probed to determine the contribution of W ions. The obtained curves are shown in Fig. \ref{lambda}. The results reveal that the double-well potential vanishes in the case of the La+N mode, indicating that the La atom is dynamically unstable on its intermediate position. A double-well potential appears in the case of the W+N mode, although the depth of the wells is slightly shallower than those of the full mode. This indicates that ferroelectricity in LaWN$_3$ mostly arises from motion of the W ions, and not the La ion displacements. This is unusual because ferroelectricity in LiNbO$_3$-type materials is associated with \emph{A}-site displacement for most oxide perovskites with tolerance factors less than 1.\cite{Inaguma2008,Xiang-PRB-2014,Wang2015PRL,Taka2015}

In the $R3c$ form of LaWN$_3$ ($t$ = 0.969), although the $A$-site cation is displaced from its centrosymmetric site, this does not give rise to its ferroelectricity. The microscopic origin of ferroelectricity in LaWN$_3$ thus appears to be similar to that in LiNbO$_3$. In LiNbO$_3$, the ferroelectric instability is driven by the hybridization of Nb-4$d$ and O-2$p$ states, which causes O anions to be displaced toward the Nb ions; displacements of O anions result in displacements of Li ions from their centrosymmetric positions.\cite{CohenPRB1996} Similarly, the driving mechanism of ferroelectricity in LaWN$_3$ is the hybridization of W-5$d$ and N-2$p$ states (see Figs. S4 and S5 in Supplementary Material), which leads to relative displacements of W cations and N anions, with the displacements of N cations pushing La ions off their centrosymmetric positions. $A$-site cations thus play a passive role in the ferroelectric transition in LaWN$_3$, as shown in Fig. \ref{lambda}, only displacements of W ions drive the ferroelectric instability of LaWN$_3$, in contrast, both Nb and Li displacements contribute to the ferroelectric instability of LiNbO$_3$.\cite{Xiang-PRB-2014}

\section{Summary}\label{sec-IV}
First-principles calculations of perovskite-type ferroelectric nitride LaWN$_3$ within the framework of density functional theory have enabled us to determine the key structural distortions controlling the ferroelectric phase transition. The $Pm\bar{3}m$ (cubic) form is unstable at 0 K and will spontaneously transform to the ground-state $R3c$ phase. By studying the relationship between structure stability and symmetry-adapted distortions, two transition pathways associated with ground-state $R3c$ LaWN$_3$, viz. $Pm\bar{3}m$$\rightarrow$$R\bar{3}c$$\rightarrow$$R3c$ and $Pm\bar{3}m$$\rightarrow$$R3m$$\rightarrow$$R3c$, have been elucidated. The strong hybridization of W-5$d$ and N-2$p$ states produces a band gap of 1.72 eV according to hybrid functional calculations. The strong interaction results in displacements of W and N ions, indirectly leading to displacement of La ions through their interaction with N ions. All these displacements are involved in optimizing the coordination environments of $A$-site and $B$-site cations, but only the displacements of $B$-site W cations along the pseudocubic [111] direction drive the ferroelectric instability in $R3c$ LaWN$_3$. The large spontaneous polarization of about 61 $\mu$C/cm$^2$ and exothermic formation enthalpy of -11.617 eV/f.u. calculated for LaWN$_3$ indicate it should be a good ferroelectric material synthesizable under appropriate conditions. Experiments to determine the optimum method and conditions for synthesizing LaWN$_3$ and to confirm its ferroelectric behavior are currently under way.
\subsection*{ACKNOWLEDGMENTS}
The computational sources have been provided by the computing center of ECNU, Nanostructures Research Laboratory of JFCC, and Chinese Tianhe-1A system of National Supercomputer Center. The authors are grateful to N. Otani, A. Konishi, K. Shitara, and Hongjian Zhao for valuable discussions.
The work at ECNU was supported by the National Basic Research Program of China (Grant No. 2014CB921104, 2013CB922301) and NSFC (Grant No. 51572085). Y.-W.F. thanks the Doctoral Mobility Scholarship at ECNU for funding his stay at JFCC. R.H. was partially supported by Natural Science Foundation of Shanghai (Grant No. 16ZR1409500)

%

\clearpage
\appendix

\renewcommand{\thefigure}{S\arabic{figure}}
\renewcommand{\figurename}{Figure}
\setcounter{figure}{0}

\title{Supplementary Material for\\ ``Lattice dynamics and ferroelectric properties of nitride perovskite LaWN$_3$''}

\author{{Yue-Wen Fang$^{1,2}$, Craig A.J. Fisher$^2$, Akihide Kuwabara$^2$, Xin-Wei Shen$^1$, Takafumi Ogawa$^2$, Hiroki Moriwake$^{2,\dag}$, Rong Huang$^{1,2}$, and Chun-Gang Duan$^{1,3,*}$}\\
{\small \em $^1 $Key Laboratory of Polar Materials and Devices, Ministry of Education, \\
Department of Electronic Engineering, East China Normal University, Shanghai 200241, China\\
$^2$Nanostructures Research Laboratory, Japan Fine Ceramics Center, Nagoya 456-8587, Japan\\
$^3$Collaborative Innovation Center of Extreme Optics, Shanxi University,
 Taiyuan, Shanxi 030006, China\\
}}

\date{\today}
\maketitle

\section{I. Structural stability of $Pnma$ and $Pna2_{1}$ structures}
\begin{figure}[htbp]
  \centering
  \includegraphics[width=0.99\linewidth]{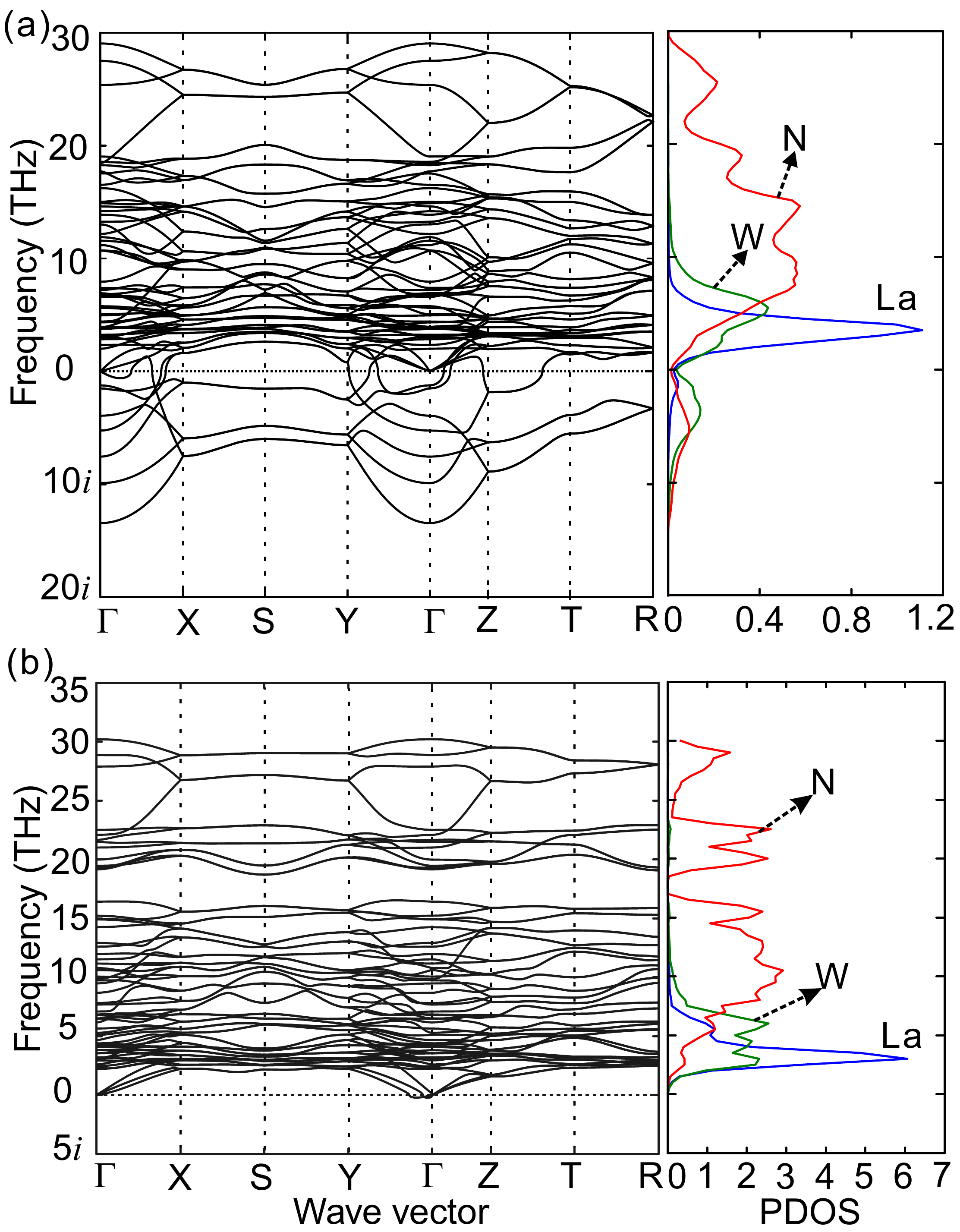}\vspace{-2pt}
  \caption{(Color online) Phonon dispersion curves and projected phonon densities of states (PDOSs) of (a) $Pnma$ and (b) $Pna2_1$ LaWN$_3$. In the PDOS plots, blue, green, and red curves represent contributions of La, W, and N atoms, respectively.}\label{FIGS1}
\end{figure}

Calculated phonon dispersion curves and projected phonon density of states (PDOS) of $Pnma$ LaWN$_3$ are shown in Fig. \ref{FIGS1}(a). Strongly unstable modes occur at the $\Gamma$ point. Freezing in these unstable modes leads to the $Pna2_1$ phase. Fig. \ref{FIGS1}(b) shows the phonon dispersion curves and PDOS of $Pna2_1$ LaWN$_3$. In this case, the unstable modes have almost disappeared at the $\Gamma$ point, implying it is more dynamically stable than the $Pnma$ phase.

\section{II. Structural phase transitions associated with $Pna2_1$ symmetry}
\begin{figure}
  \centering
  \includegraphics[width=0.99\linewidth]{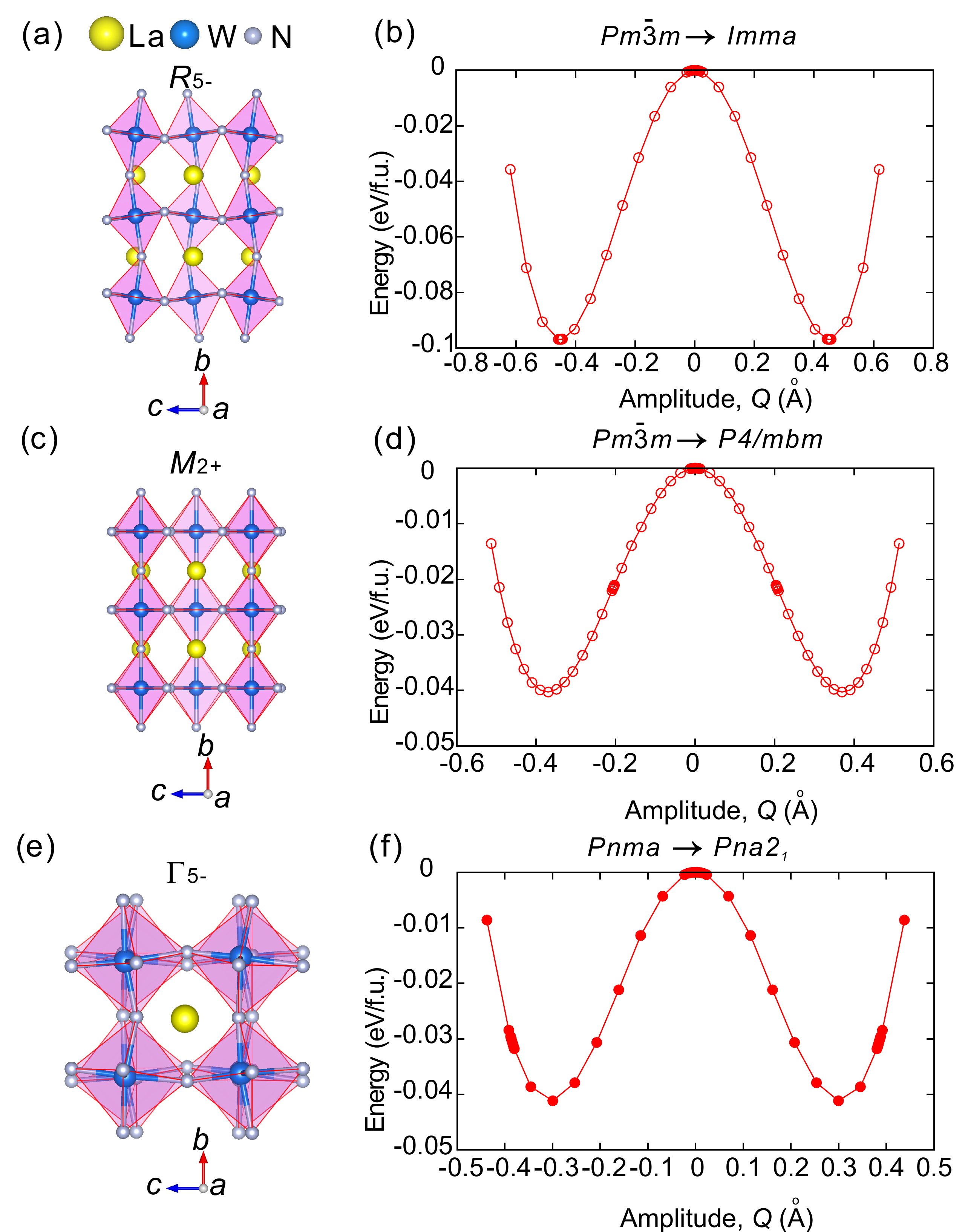}\\
  \caption{(Color online) Schematics of (a) the $Imma$ reference structure with only the $R_{5-}$ mode frozen, (b) the $P4/mbm$ reference structure with only the $M_{2+}${} mode frozen, and (c) the $Pna2_1$ reference structure with the $\Gamma_{5-}$ mode frozen.
  (d), (e) and (f) show the energy-amplitude curves for $Pm\bar{3}m$$\rightarrow$$Imma$, $Pm\bar{3}m$$\rightarrow$$P4/mbm$, and $Pnma$$\rightarrow$$Pna21$ transitions, respectively. Energies in (d) and (e) are relative to that of the $Pm\bar{3}m$ structure, while those in (e)
  the $Pnma$ structure.}
  \label{FIGS2:transition-pna21} %
\end{figure}

The pathway $Pm\bar{3}m$$\rightarrow$$Pnma$$\rightarrow$$Pna2_1$ was investigated by calculating the energy per formula unit (f.u.) as a function of mode amplitude. The transition from the $Pm\bar{3}m$ phase to $Pnma$ phase involves a combination of $R_{5-}$ and $\Gamma_{2+}$ distortion modes; hence the $Pm\bar{3}m$$\rightarrow$$Pnma$ transition is a combination of $Pm\bar{3}m$$\rightarrow$$Imma$ and $Pm\bar{3}m$$\rightarrow$$P4/mbm$ transitions.
The $Imma$, $P4/mbm$, and $Pna2_1$ reference structures are shown in Figs. \ref{FIGS2:transition-pna21}(a), (b), and (c), respectively.
Energy-amplitude curves for $Pm\bar{3}m$$\rightarrow$$Imma$, $Pm\bar{3}m$$\rightarrow$$P4/mbm$, and $Pnma$$\rightarrow$$Pna2_1$ transitions are shown in Figs. \ref{FIGS2:transition-pna21}(d), (e), and (f), respectively.
For the transition $Pnma$$\rightarrow$$Pna2_1$, the symmetry-adapted mode is mainly composed of the $\Gamma_{5-}$ mode.

\section{III. Electronic properties and ferroelectric polarization of the $Pna2_1$ phase}
The band gap of $Pna2_1$ LaWN$_3$ obtained from HSE hybrid calculations is about 1.25 eV. The ferroelectric polarization was calculated to be about 20 $\mu$C/cm$^2$.

\section{IV. Effect of semi-core W $p$ states on ferroelectric instability}
\begin{figure}
  \centering
  \includegraphics[width=0.99\linewidth]{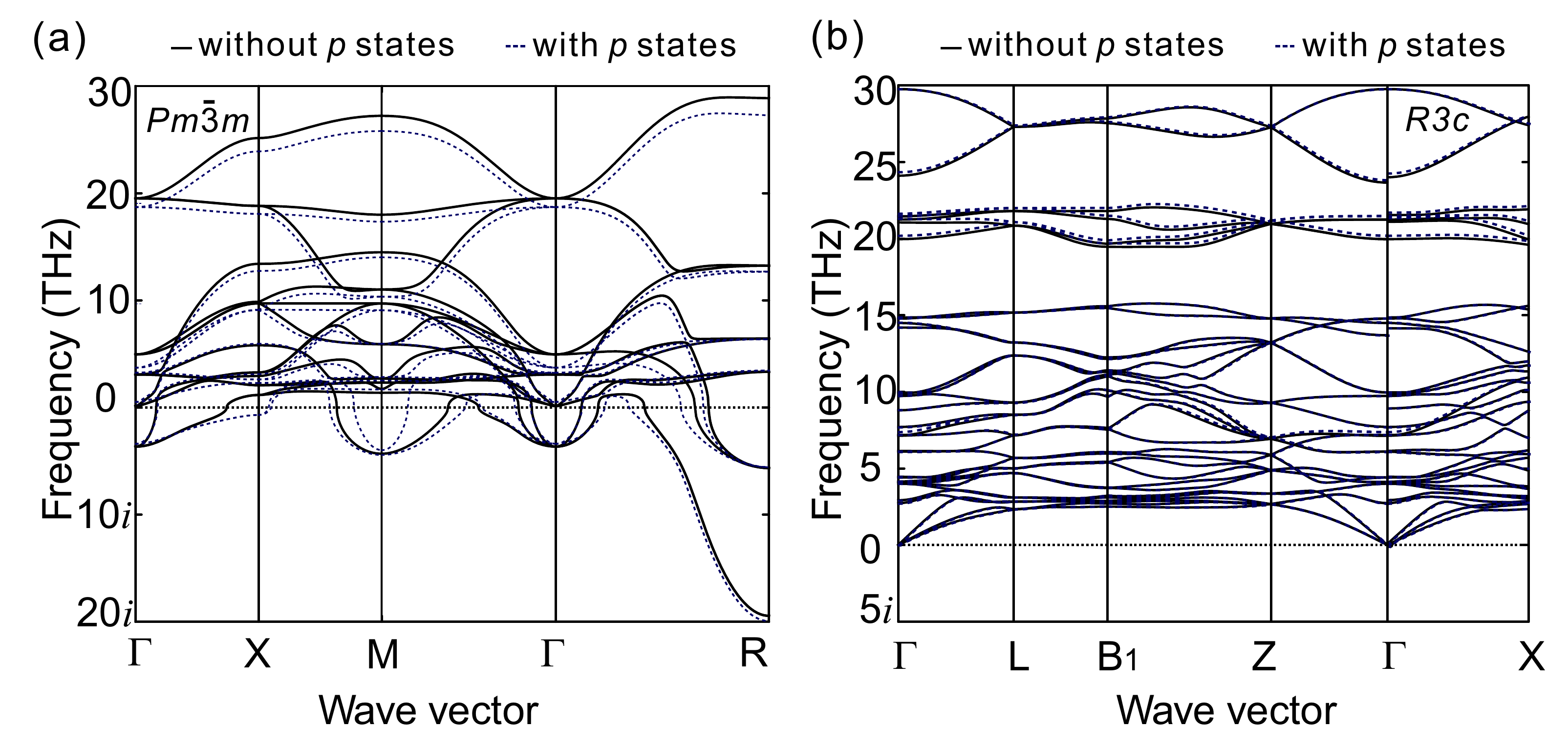}\\
  \caption{(Color online) Phonon dispersion curves of (a) $Pm\bar{3}m$ and (b) ferroelectric $R3c$ LaWN$_3$. Phonon dispersion curves calculated with (without) semi-core W $p$ states are plotted as black solid (blue dashed) lines.}
  \label{FIGS3:pstates} %
\end{figure}

Figure \ref{FIGS3:pstates} compares the phonon dispersion curves of (a) $Pm\bar{3}m$ and (b) $R3c$ LaWN$_3$.
For $Pm\bar{3}m$ LaWN$_3$, pseudopotentials with and without semi-core W $p$ states included both reproduce the unstable modes at the $\Gamma$ point. Furthermore, the phonon dispersions along the high-symmetry line ($\Gamma$ $\rightarrow$ $L$ $\rightarrow$ $B_1$ $\rightarrow$ $Z$ $\rightarrow$ $\Gamma$ $\rightarrow$ $X$) of $R3c$ LaWN$_3$ are almost identical.
These results indicate that semi-core W $p$ states have little influence on the strong ferroelectric instability of LaWN$_3$.

Ferroelectric polarization values calculated using pseudopotentials with and without semi-core W $p$ states were also the same (61 $\mu$C/cm$^2$).

Therefore, the psudopotentials used in our main text are adequate for reproducing the strong ferroelectric instability of LaWN$_3$.

\section{V. The influence of hybridization between W-5$d$ and N-2$p$ on the ferroelectric polarization}
Using the orbital selective external potential (OSEP) method,\cite{Wan2010,FANG2015} we investigated the effects of hybridization of unoccupied W-5$d$ and N-2$p$ orbitals on the ferroelectric polarization of $R3c$ LaWN$_3$. 
This method has previously been implemented and used to study the microscopic mechanism of ferroelectricity in oxide perovskite.\cite{FANG2015}

\begin{figure}
  \centering
  \includegraphics[width=0.99\linewidth]{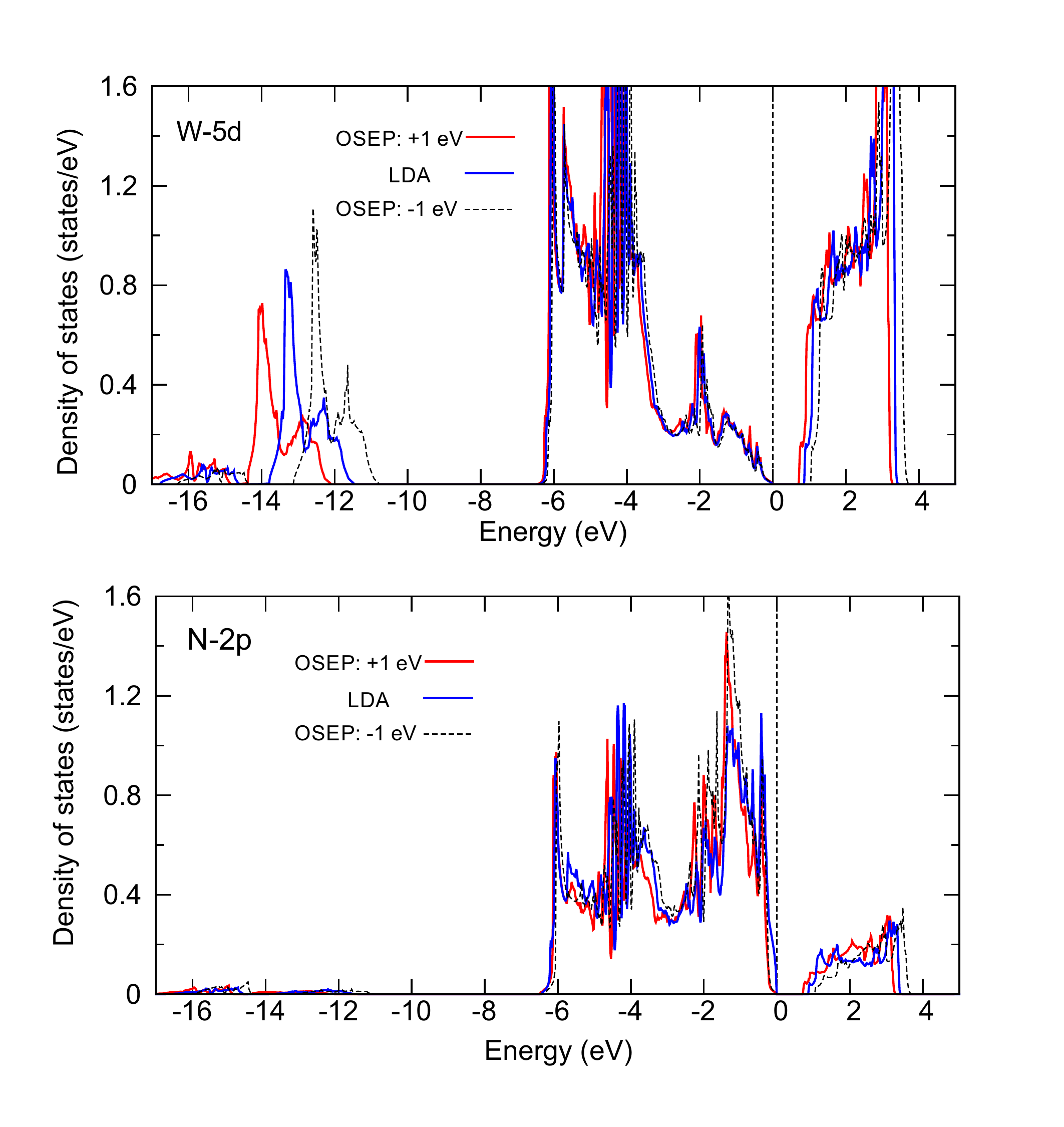}\\
  \caption{(Color online) Partial densities of states of W-5$d$ and N-2$p$ orbitals calculated by using OSEP and LDA methods. In the OSEP calculations, an external field of $\pm$1 eV was applied to the N-2$p$ orbitals.}
  \label{FIGS4:hybridization} %
\end{figure}

\begin{figure}
  \centering
  \includegraphics[width=0.99\linewidth]{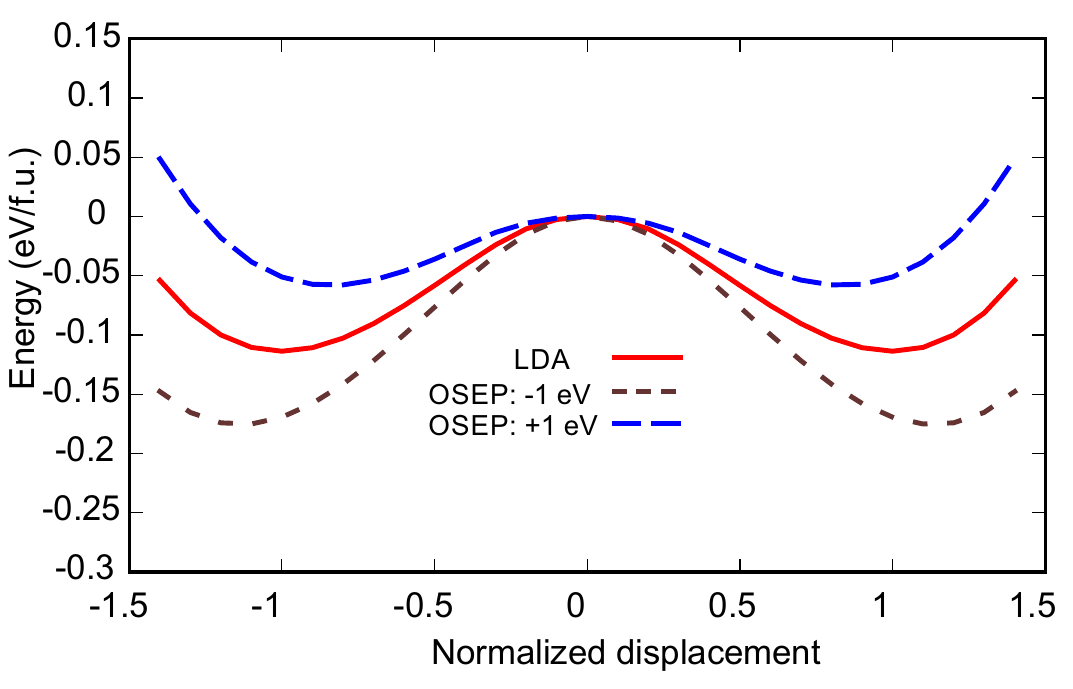}\\
  \caption{(Color online) Energy as a function of normalized displacement obtained using OSEP and LDA methods. In the case of LDA calculations, a normalized displacement of 1 corresponds to the ferroelectric $R3c$ structure, while 0 displacement corresponds to the paraelectric $R\bar{3}c$ structure.}
  \label{FIGS5:OSEPdoublewell} %
\end{figure}

External fields were applied to N-2$p$ orbitals to modulate the strength of cross-gap hybridization between N-2$p$ and W-5$p$ states. Figure \ref{FIGS4:hybridization} compares the partial densities of states of W-5$d$ and N-2$p$ orbitals calculated by OSEP and LDA methods. The results show that cross-gap hybridization between N-2$p$ and W-5$d$ states is enhanced (decreased) by applying a -1 eV (+1 eV) field to N-2$p$ orbitals.
The corresponding double-well shaped energy-amplitude curves are shown in Fig. \ref{FIGS5:OSEPdoublewell}.
The curves are affected strongly when an electric field is applied. A +1 eV field on N-2$p$ orbitals makes the double-well shallower, while a -1 eV field makes the double-well deeper.
In the former case, the calculated ferroelectric polarization is 59.6 $\mu$C/cm$^2$ ($<$ 61 $\mu$C/cm$^2$ in LDA), while in the latter case the polarization is 61.4 $\mu$C/cm$^2$ ($>$ 61 $\mu$C/cm$^2$ in LDA).
These results confirm that the cross-gap hybridization between W-5$d$ and N-2$p$ states is the driving force for ferroelectricity in LaWN$_3$.

\clearpage
\newpage
%
%
%
%


\end{document}